\documentclass[prd,aps,twocolumn,amsfonts,showpacs,superscriptaddress,preprintnumbers,nofootinbib]{revtex4-2}
 
\usepackage{graphicx}
\usepackage{xcolor}
\usepackage{rotating}
\usepackage{bm,amsmath,amssymb}
\usepackage[utf8]{inputenc}
\usepackage{mathabx}
\usepackage{comment}
\usepackage{footmisc}
\usepackage{diagbox}
\usepackage{amsmath}
\usepackage{cancel}
\usepackage{physics}
\usepackage{dcolumn}
\usepackage{bm}
\usepackage{hyperref}
\usepackage{colortbl}
\definecolor{ddred}{RGB}{198,31,31}
\definecolor{ddgreen}{RGB}{0,160,77}
\hypersetup{
    colorlinks=true,
    linkcolor=ddred,     
    urlcolor=ddred,
    citecolor=ddgreen}

\begin{document}


\title{Quasinormal modes as exterior probes of black hole interiors in our Universe}

\author{Sa\v{s}o Grozdanov}
\affiliation{Higgs Centre for Theoretical Physics, University of Edinburgh, Edinburgh, EH8 9YL, Scotland, 
}
\affiliation{Faculty of Mathematics and Physics, University of Ljubljana, Jadranska ulica 19, SI-1000 Ljubljana, Slovenia
}

\author{Vita Movrin}
\affiliation{Faculty of Mathematics and Physics, University of Ljubljana, Jadranska ulica 19, SI-1000 Ljubljana, Slovenia
}

\author{Samuel Valach}
\affiliation{Faculty of Mathematics and Physics, University of Ljubljana, Jadranska ulica 19, SI-1000 Ljubljana, Slovenia
}

\begin{abstract}
We consider a scenario in which a black hole is enclosed in a reflecting timelike cavity and linearly perturbed. For the resulting quasinormal spectra of gravitational, electromagnetic, and other perturbations, in conjunction with our Ref.~\cite{Grozdanov:2026ktq}, we establish a thermal product formula and derive a universal asymptotic expression for high quasinormal overtones. Both results hold in asymptotically flat, de Sitter, and Anti-de Sitter spacetimes (known previously). This spectral signature directly encodes information about the black hole interior and its singularity, and is, at least in principle, accessible to an observer outside the event horizon. We then propose a concrete observational setup in which accreting plasma plays the role of a partially transmissive mirror. For small, hypothetical, primordial black holes, we argue that one may thereby be able to extract concrete information about the black hole interior from electromagnetic waves in the radiowave spectrum.
\end{abstract}

\maketitle


{\bf Introduction and motivation.---}Due to advances in observational astronomy, the existence of astrophysical black holes now seems firmly established. Supermassive black holes at the centres of galaxies visibly influence the trajectories of nearby stars \cite{Schodel:2002vg,Ghez:2008ms}. Numerical-relativity-based waveform models successfully describe the measured gravitational waves from binary mergers of black holes with masses of a few to several tens of solar masses \cite{LIGOScientific:2016aoc}. Moreover, very-long-baseline radio interferometry has produced horizon-scale images of the emission surrounding nearby supermassive black holes \cite{EventHorizonTelescope:2019dse}. Smaller primordial black holes (PBHs) are also of considerable theoretical interest (e.g., as possible dark matter candidates), however, their existence has not been observationally confirmed (see e.g., Ref.~\cite{Carr:2021bzv}).

A black hole's semi-classical description is universally characterised by a small number of distinct features: namely, the outer edge of the black hole or the event horizon, a potential inner (Cauchy) horizon, and a timelike or spacelike singularity. The mathematical properties of event horizons imply that black holes behave as thermodynamic bodies. Those thermal properties associated with the Hawking radiation are, at least in principle, observable to outside observers \cite{Hawking:1975vcx}. The physics of various quantum observables outside the black hole is thereby expected to behave as if in a thermal state $\hat\rho \sim e^{-\beta \hat H}$, where $\beta = 1/T$ is the inverse (Hawking) temperature. 

Devising external probes of black hole interiors that could even theoretically `probe' the physics near singularities has been, unsurprisingly, a much more difficult task. While a number of such probes have been constructed in the context of the AdS/CFT correspondence, e.g., bouncing geodesics \cite{Fidkowski:2003nf,Festuccia:2005pi,Ceplak:2024bja} and measures of complexity \cite{Stanford:2014jda,Brown:2015bva}, these probes typically involve an analytic continuation to a two-sided black hole, with the relevant bulk objects connecting two spacelike-separated exterior regions through the black hole interior.

Building on such past analyses of black holes in Anti-de Sitter space (AdS), in this paper, we present a setup that should enable an external observer, at least in principle, to probe the interiors of black holes resembling those in our Universe using light spectroscopy. This is made possible by a universal relation governing the spacing of highly damped quasinormal modes (QNMs) of black holes enclosed in a cavity, which we establish for spacetimes with arbitrary cosmological constant $\Lambda$. As we demonstrate, such a cavity could be provided by a highly reflecting accreting plasma. While this claim may at first seem highly counter-intuitive, analogous behaviour of QNMs in AdS was known before. In computing spectroscopic predictions for the QNM spectrum, one assumes a static black hole with a spacetime geometry that remains analytic up to the singularity. As one solves the relevant differential equations between the horizon and an asymptotic observer, the analytic structure of their solutions, inherited from the singular points of the equations and from their monodromy structure in the complexified spacetime, becomes imprinted on the spectrum. In our case, this imprint of the black hole interior proves to be universal also for more realistic asymptotically flat and de Sitter (dS) black holes. This therefore offers a potential future observational window into black hole interiors.

{\bf Universal asymptotic spacing of QNMs in a reflecting cavity.---}Consider a massive, non-rotating Schwarzschild black hole in 4$d$ Einstein gravity with an arbitrary $\Lambda$. We study its linearised perturbations, be it scalar or electromagnetic waves expanded around vanishing background values, or gravitational perturbations. We claim that all such spectra, when placed in a reflecting cavity, follow a universal relation that sets the spacing between asymptotic, relaxing QNMs. As shown in Appendix~\ref{App:CTPF} and Ref.~\cite{Grozdanov:2026ktq}, this result follows from the ``cavity thermal product formula'', which is a meromorphic factorisation formula that fixes the frequency-dependent structure of two-point correlators. This generalises the result for $\Lambda < 0$ from Ref.~\cite{dodelson2023thermal}. 

To show this statement, we first decompose the fluctuations into decoupled gauge-invariant master equations. Then, we consider the maximal analytic extension of the Schwarzschild background, which allows us to define the thermal two-sided correlator $G_{12}(\omega,z;z')$ of any of the relevant gauge-invariant master fields $\psi$. To simplify notation, we suppress (non-radial) spatial dependence and its Fourier conjugates. In terms of the standard Wightman correlator, $G_{12}(t) = G_W(t-i\beta/2)$, and, in momentum space, in terms of the retarded correlator $G(\omega)$, $G_{12}(\omega) = (G(\omega)-G(-\omega))/ 2i\sinh(\beta\omega/2)$. Since $G_{12}$ vanishes at the cavity wall, chosen to lie at $z=0$, we instead define a ``boundary correlator''
\begin{equation}
\label{eq:wall-correlator-definition}
    G_{12}^{\partial}(\omega)
    \propto
    \left.
    \partial_z\partial_{z'}G_{12}(\omega,z;z')
    \right|_{z=z'=0}.
\end{equation}
Equivalently, if the solution ingoing at the horizon has a regular near-wall
expansion
$h_+(\omega,z)=A(\omega)+B(\omega)z+\mathcal O(z^2)$, then the retarded (response) $G^{\partial}(\omega) \propto B(\omega)/A(\omega)$, which is reminiscent of the standard holographic correlator structure in AdS. Under the analyticity and growth assumptions established in Ref.~\cite{Grozdanov:2026ktq}, the inverse of
$G_{12}^{\partial}(\omega)$ is an entire function of order one whose
zeros are the cavity quasinormal frequencies. Since the poles of $G_{12}^{\partial}(\omega)$ occur in quartets
built from the retarded QNMs and their reflections,
$(\omega_n,-\omega_n,\omega_n^*,-\omega_n^*)$, the
Weierstrass--Hadamard theorem gives the cavity thermal product formula:
\begin{equation}
\label{eq: CTPF}
    G_{12}^{\partial}(\omega)
    = \frac{G_{12}^{\partial}(0)}
    {\displaystyle\prod_{n=1}^{\infty}
    \left(1-\frac{\omega^2}{\omega_n^2}\right)
    \left(1-\frac{\omega^2}{(\omega_n^*)^2}\right)} ,
\end{equation}
which expresses the meromorphic structure typical of thermal (holographic) black hole correlators \cite{Kovtun:2005ev,Hartnoll:2005ju,Grozdanov:2016vgg,Grozdanov:2018gfx,dodelson2023thermal}. 

The large-$\omega$ behaviour of \eqref{eq: CTPF} then relates the asymptotic QNM spectrum to singularities of the correlator in complex time. By the local Hadamard form and the propagation of singularities theorem (see our Refs.~\cite{Grozdanov:2026cut,Grozdanov:2026ktq}), the first nontrivial singularity of the retarded $G^\partial(t)$ is associated with the bouncing null geodesic at complex time $t_*$, fixing the asymptotic behaviour of QNMs:
\begin{equation}
\label{main_eq}
    \omega_n\sim\frac{2\pi n}{t_*},
    \qquad n\to\infty .
\end{equation}
For details of the derivation, see Appendix~\ref{App:CTPF} and Ref.~\cite{Grozdanov:2026ktq}. 


{\bf Spectrum as exterior signature of the black hole singularity.---}We now discuss the meaning of $t_*$, which directly encodes the information about the black hole interior. Concretely, this bouncing time is twice the (analytically continued) Schwarzschild time it takes for light to reach the singularity at $r=0$, starting from the cavity wall at $r_i$ (or $z=0$). For a black hole with a metric given in terms of the ``Schwarzschild coordinates'',
\begin{equation}
    \dd s^2=-f(r)c^2\dd t^2+\frac{\dd r^2}{f(r)}+r^2\dd\Omega_2^{2},
\end{equation}
where $c$ is the speed of light, $t_*$ is computed as
\begin{equation}\label{e.rubber_ducks_look_so_real}
    t_*=\pm\frac{2}{c}\int_{r_i}^0\frac{\dd r}{f(r)},
\end{equation}
where the sign is fixed by requiring that $\Re t_*>0$ for $r_i\to\infty$. Importantly, in order for $t_*$ to be connected to the momentum space singularities via Eq.~\eqref{main_eq}, it needs to be related to a certain (``bouncing'') geodesic. See Ref.~\cite{Grozdanov:2026cut} for more details. 

For 4$d$ Schwarzschild black holes studied here,
\begin{equation}
    f(r)=1-\frac{2GM}{c^2r}-\frac{\Lambda}{3}r^2,
\end{equation}
where $M$ is the mass of the black hole and $G$ is the gravitational constant. For asymptotically flat ($\Lambda=0$), AdS ($\Lambda<0$) and dS ($\Lambda>0$) black holes, Eq.~\eqref{e.rubber_ducks_look_so_real} yields
\begin{align}
    t_*^{\Lambda=0}&=\frac{2}{c}\left[r_i+\frac{2GM}{c^2}\ln\left(1-\frac{c^2r_i}{2GM}\right)\right],\label{e.tstar_zero}\\
    t_*^{\Lambda<0}&=\frac{1}{c}\sum_{k=1}^3\frac{2r_k}{\Lambda r_k^2-1}\ln\left(\frac{r_k}{r_k-r_i}\right),\label{e.tstar_minus}\\
    t_*^{\Lambda>0}&=\frac{\sqrt{f\!\left(\!\sqrt[3]{\frac{3GM}{c^2\Lambda}}\right)}}{c}\sum_{k=1}^3\frac{2r_k}{\Lambda r_k^2-1}\ln\left(\frac{r_k}{r_k-r_i}\right),\label{e.tstar_plus}
\end{align}
where $r_k$ are the three roots of $f(r)=0$. For $\Lambda<0$, only one is real and positive, and it corresponds to the location of the event horizon $r_b$. For $\Lambda>0$, there is a second positive real root at the cosmological horizons $r_c$, and we used a different overall normalisation that measures time with respect to the stationary sphere observer at $f'(r)=0$ (for details, see Ref.~\cite{Grozdanov:2026ktq}).

Finally, note that if the anchoring point $r_i$ at the cavity wall lies outside the black hole horizon, the bouncing times \eqref{e.tstar_zero}--\eqref{e.tstar_plus} generically acquire a constant imaginary part $i\beta/2$, making $t_*$ complex. This is consistent with the expectation that the corresponding cavity QNMs have $\Im \omega < 0$ for reasons of stability.

The validity of Eq.~\eqref{main_eq} for different $\Lambda$ was checked numerically by computing the cavity QNMs for scalar, electromagnetic (EM) and gravitational perturbations in our Ref.~\cite{Grozdanov:2026ktq}. Here, however, we now turn to discussing only the EM perturbations, which will be of relevance to the proposed phenomenological setup below. In 4$d$ asymptotically-flat Schwarzschild spacetime, both the parity-even and parity-odd channels of EM perturbations described by master fields $\psi_\ell^\pm$ obey the same master equation:
\begin{equation}
\left[\frac{d^2}{dr_*^2} +\frac{\omega^2}{c^2} -f(r)\frac{\ell(\ell+1)}{r^2} \right]\psi_\ell^\pm=0,
\end{equation}
where $dr_*=dr/f(r)$. A perfectly reflecting conducting wall at $r=r_i$ is implemented by requiring a vanishing energy flux of the Maxwell field through the wall, discussed in Refs.~\cite{Wang:2015goa,Lei:2021kqv}. Importantly, however, this condition does not uniquely fix the boundary conditions for the Maxwell field. We need to impose a stronger, \textit{perfect electric conductor} (PEC) condition, whereby the tangential components of the electric 
and the normal components of the magnetic fields vanish at the timelike cavity wall \cite{Brito:2015oca}. In terms of the master fields, then, 
\begin{equation}
    \psi_\ell^-(r_i)=0,
    \qquad
    \partial_r\psi_\ell^+(r_i)=0 .
\end{equation}
Thus, the odd channel obeys a Dirichlet condition, while the even channel obeys a Neumann condition. Since the cavity thermal product formula \eqref{eq: CTPF} only requires a regular, $\omega$-independent reflecting condition at the wall, both channels have the same asymptotic spacing,
\begin{equation}
    \omega_{n+1}^{\pm}-\omega_n^{\pm}
    \simeq \frac{2\pi}{t_*(r_i)} .
\end{equation}


{\bf A hypothetical measurement of $t_*$ for black holes in our Universe.---}The result in Eq.~\eqref{main_eq} universally applies to a black hole in a cavity with any cosmological constant  and relates a simple geometric quantity that probes its interior with the QNM spectrum of different types of fluctuating linearised fields. Moreover, as shown in Ref.~\cite{Grozdanov:2026ktq}, even the spacing between long-lived (low-$n$) QNMs is very well approximated by this relation. 

The next natural question is therefore whether $t_*$ could be directly inferred for a realistic black hole from the measurements of its QNM ringdowns. While the gravitational wave detections triggered by black hole mergers have been able to detect QNMs beyond the first one \cite{Isi:2019aib,LIGOScientific:2020tif,Ghosh:2021mrv,Capano:2021etf}, these spectra are controlled by outgoing boundary conditions at infinity. Moreover, devising a mechanism that would reflect gravitational waves off the cavity wall seems rather unrealistic. Instead, we focus on electromagnetic radiation, which can be reflected off a timelike boundary by a ``mirror''. Without the need to build a Dyson sphere, the black hole accretion disk can act as a natural (at least partially) reflecting mirror of light. For many realistic black holes, such as the Sagittarius A* (or Sgr A*; the galactic centre of the Milky Way) and M87*, the accretion disc, composed of a highly energetic plasma, covers roughly 50--80\% of the solid angle ``seen'' from the center of the black hole \cite{Narayan:1994is,Yuan:2014gma,EventHorizonTelescope:2022urf}. It reflects well the EM frequencies below its plasma (angular) frequency, which can be approximated to be about $\omega_p\sim 10^8 s^{-1}$, or in terms of frequency, $\nu_p = \omega_p / 2\pi \sim 10 \, \rm{MHz} $  \cite{rybicki1979radiative,Bower:2019rsg}.  

\begin{figure}[h!]
\centering
\includegraphics[scale=0.64]{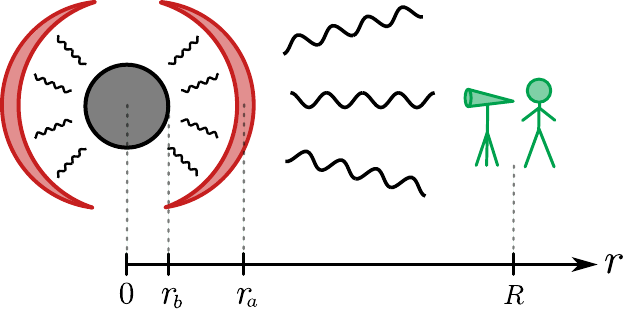}
\caption{Setup of a hypothetical experiment for measuring electromagnetic QNMs that pass through the accretion disk.}
\label{f.experiment}
\end{figure}

We now consider the following hypothetical and idealised, but (approximately) naturally occurring scenario as our observational setup. Neglecting rotation, a black hole with its centre at $r=0$ and event horizon at $r=r_b$ is enclosed in an approximately spherical accretion disk at $r=r_a$ characterised by the plasma frequency $\omega_p$ that reflects EM radiation. Some of that light is transmitted through the plasma, of which the spectrum is then measured by an (asymptotic) observer at $r=R$ and used to infer $t_*$. The setup is depicted in Figure~\ref{f.experiment}.

Since the black hole is now not surrounded by a perfectly reflecting mirror, its QNM spectrum will differ from the one to which the cavity thermal product formula \eqref{eq: CTPF} and Eq.~\eqref{main_eq} strictly apply. However, as we show below, the changes to the spectrum may be very small, and so long as the transmission of radiation through the plasma remains non-negligible, this may enable a potential measurement of $t_*$ from the low-$n$ QNMs. 

Before modeling such a transparent accretion disk, we consider the scales of frequencies that enter Eq.~\eqref{main_eq} and $t_*$ for (Schwarzschild) black holes with the masses of the following three black holes: the supermassive Sgr A* with mass $M\sim4.3\times 10^6 M_\odot$ (where $M_\odot$ is the mass of the Sun), the closest known black hole to Earth called Gaia BH1 with mass $M\sim10M_\odot$, and a hypothetical small few-Earth-mass primordial black hole (PBH). One such PBH has been suggested in our solar system as an explanation for the clustering of orbits of the Kuiper belt objects \cite{Scholtz:2019csj,Witten:2020ifl}. In terms of the Earth's mass $M_{\Earth}$, its mass would be on the order of $M\sim10M_{\Earth}$. This PBH (which we call sPBH) is simply considered as a motivating example of any, hypothetical, small PBH. Note that both Gaia BH1 and sPBH (if it exists) are currently in a dormant phase, i.e., they are not actively ``feeding'' on surrounding matter. Nevertheless, in all considered cases, we assume the accretion disc to start at $r_a \gtrsim 3r_b$, i.e.,~at or close to the innermost stable circular orbit (ISCO) of the Schwarzschild black hole.

\begin{figure}[h!]
\centering
\includegraphics[scale=1.1]{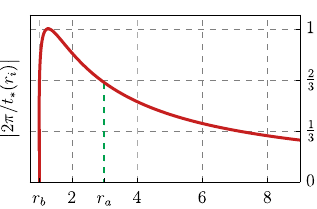}
\caption{The asymptotic cavity QNM spacing plotted as a function of the anchoring distance $r_i$, in units of $r_b=1$.}
\label{f.omega}
\end{figure}

In Figure~\ref{f.omega}, setting $\Lambda = 0$ and using Eqs.~\eqref{main_eq} and \eqref{e.tstar_zero}, we show the dependence of $|2\pi/t_*|$ on the location of the timelike anchoring surface of the geodesic, i.e., location of the reflecting wall. Its value increases near the event horizon, remains appreciable in the accretion disk region near the ISCO, and then decreases.  

\begin{figure}[ht!]
\centering
\includegraphics[scale=1]{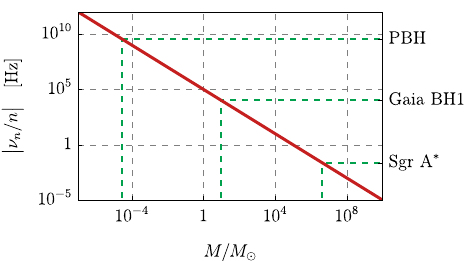}
\caption{Log-log plot of the mass-dependent frequency spacing $\abs{\nu_n/n}$ for $r_i=r_a=3r_b$, as predicted by Eq.~\eqref{main_eq}.}
\label{f.mass}
\end{figure}

With the choice of $r_i = 3 r_b$, we find the following predictions for the cavity QNM frequencies of black holes with the three corresponding masses:
\begin{align}
    \text{Sgr A*:}\quad \nu_n/n&\approx(1.9-1.6\,i)\,\text{mHz},\\
    \text{Gaia BH1:}\quad \nu_n/n&\approx(0.8-0.7\,i)\,\text{kHz},\\
    \text{sPBH:}\quad \nu_n/n&\approx(0.3-0.2\,i)\,\text{GHz},\label{e.LIL}
\end{align}
where we neglect any effect of cosmological redshift. Indeed, for black holes within the Milky way, one expects to measure approximately the same QNMs $\omega_n = 2\pi \nu_n$, irrespectively of the distance $R$ of the observer from the accretion disc. Since the very low frequencies associated with supermassive black holes are particularly challenging observationally, our hypothetical experiment seems to be better suited to probing smaller black holes, especially PBHs. See Figure~\ref{f.mass}.


{\bf The QNM spectrum transmitted through an accreting plasma.---}We now relax the condition of a perfectly reflecting cavity and model the setup depicted in Figure~\ref{f.experiment} with $\Lambda=0$. In particular, we consider axial EM perturbations with the radial master equation
\begin{equation}\label{e.apple_on_tree}
    \frac{d^2\psi_\ell}{dr_*^2}  +\left[\frac{\omega^2}{c^2}-V_\ell(r,\omega)\right]\psi_\ell=0.
\end{equation}
A simple effective potential, using a factorisation between the geometry and the plasma potential, can be chosen as  
\begin{equation}\label{e.rozok}
    V_\ell(r,\omega)
    =f(r)\left[
    \frac{\ell(\ell+1)}{r^2}
    +\left(\frac{\omega_p^2}{c^2}-i\frac{\omega\gamma}{c^2}\right)W(r)
    \right],
\end{equation}
where we model the plasma shell by taking 
\begin{equation}\label{e.apple_farmer}
    W(r) = 
    \frac{1}{4}
    \left[1+\tanh\frac{r-r_a}{\delta} \right] \left[1-\tanh\frac{r-r_a-L}{\delta}
    \right].
\end{equation}
Here, $\omega_p$ is the maximum plasma frequency, while $\gamma\geq0$ accounts for absorption within the shell. The parameters $r_a$, $L$, and $\delta$ denote its inner radius, thickness, and edge width, respectively. In the highly reflective limit, the inner edge $r_a$ acts as the effective cavity wall, so the reflective-wall prediction is obtained by setting $r_i=r_a$. 

\begin{figure*}[ht!]
    \centering
    \includegraphics[width=0.9\textwidth]
        {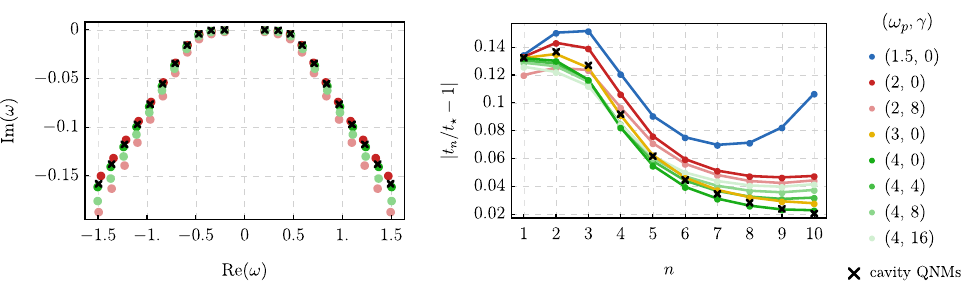}

    \caption{QNMs of axial EM perturbations with $\ell=1$. With $r_b=1$, the cavity wall and the inner edge of the plasma shell are located at $r_i=r_a=20$. We set $(L,\delta)=(5,0.5)$ and show results for several values of $\omega_p$ and $\gamma$. The transparent plasma QNMs are obtained using the pseudospectral method \cite{jansen2017overdamped}. Black crosses denote the reflecting cavity QNMs obtained with the shooting method. The right panel shows the relative deviation of the observed spacing from the prediction of Eq.~\eqref{main_eq}.}
    \label{f.plasma_qnms}
\end{figure*}

The QNMs are then computed by imposing ingoing boundary conditions at the horizon and purely outgoing conditions at infinity (the observer). By defining
\begin{equation}
\label{eq:tn}
    t_n \equiv \frac{2\pi}{\omega_{n+1}-\omega_n},
\end{equation}
as shown in Figure~\ref{f.plasma_qnms}, for appropriately chosen parameters (note the plasma location at $r_a = 20 r_b$), the prediction \eqref{main_eq} can remain approximately realised even for reasonably low-$n$ QNMs with frequencies below the effective plasma barrier, $t_n \approx t_*(r_a)$. As anticipated, increasing $\omega_p$ raises the plasma barrier and makes the
shell more reflective, producing a broader interval of cavity-like modes. The right panel of Figure~\ref{f.plasma_qnms} shows that the relative error decreases with small $n$, and that larger $\omega_p$ extends the range over which the plasma-shell spacings approach the perfect reflecting cavity prediction. The modes in this regime also lie close to the cavity frequencies in the complex $\omega$ plane (see the left panel of Figure~\ref{f.plasma_qnms}). The small $\omega_p=1.5$ (in natural units) case illustrates the breakdown of the reflecting cavity description underlying the cavity result above. Its first few spacings move towards the predicted value, but the relative error begins to increase again when the real parts of the frequencies approach the effective plasma cut-off. This introduces a natural lower bound on the plasma frequency $f(r_a)\nu^2_p>\nu^2$. At fixed $\omega_p$, increasing $\gamma$ moves the plasma-shell QNMs deeper into the lower half-plane.

The next important question pertains to the amount of radiation that reaches the observer, which is encoded in the transmission coefficient. Assuming, for simplicity, that $\gamma=0$, that $\omega_p\gg\omega$, and replacing the smooth shell \eqref{e.apple_farmer} by a simple step function potential
\begin{equation}\label{e.rectangle_is_just_a_sexy_square}
    W(r)=\Theta(r-r_a)\Theta(r_a+L-r),
\end{equation}
we can compute the transmission coefficient by employing standard WKB techniques, finding (see Appendix \ref{a.WKBB})
\begin{align}
T&\approx \frac{4\omega^2\,{\rm csch}^2\left[S(r_a+L) - S(r_a)\right]}{\omega_p^2 \sqrt{\left(1 - \frac{r_b}{r_a}\right)\left(1 - \frac{r_b}{r_a+L}\right)}},\label{e.hamster}\\
\! S(r)&=\frac{\omega_p}{c} \left[ \sqrt{r(r-r_b)} + r_b \ln\left(\sqrt{r} + \sqrt{r-r_b}\right) \right].\label{e.hamsters_friend}
\end{align}
Assuming a thin shell $L \ll r_b$, with $r_a = 3 r_b$, then gives
\begin{equation}
    T\approx6\frac{\omega^2}{\omega_p^2}{\rm csch}^2\Big(\sqrt{\frac32}\frac{L\omega_p}{c}\Big).
\end{equation}
To avoid exponential suppression, $L\omega_p/c$ must not be too large. This is either achieved by small black holes (such as a PBH), or a small accreting plasma frequency (e.g., due to a small number density of electrons). In addition, $\omega/\omega_p$ must not be too small. Requiring, e.g., $T\geq0.1\%$, sets an upper bound on $\omega_p$ (and $\nu_p$). For a black hole with the mass of the sPBH with $L=r_b/10$ and the scale of frequencies set by \eqref{e.LIL}, we require $\nu_p\lesssim10^{10}$Hz. Together with an approximate lower bound $f(r_a)\nu^2_p>\nu^2$, this defines a window of $\nu_p\in[10^8\text{Hz},10^{10}\text{Hz}]$, for which the experiment could, in principle, be performed. This is a standard range of frequencies accessible to present-day radio astronomy observations. 

{\bf Discussion and future directions.---}This work (along with our \cite{Grozdanov:2026ktq}) has proven the thermal product formula for the spectra of black holes in a reflecting cavity in spacetimes with any cosmological constant $\Lambda$. This gave rise to the asymptotic QNM spacing expression Eq.~\eqref{main_eq} expressed in terms of $t_*$, previously known in AdS spaces with $\Lambda < 0$ \cite{Fidkowski:2003nf,Festuccia:2005pi,dodelson2023thermal}, which turns the classical singularity and the properties of the interior geometry into a spectroscopic prediction. Its details depend on the charge of the black hole \cite{Ceplak:2025dds,AliAhmad:2026wem,futureNEJC}, the value of $\Lambda$ \cite{Arnaudo:2026tcy,Grozdanov:2026ktq}, higher-derivative (or stringy) corrections \cite{Dodelson:2025jff,futureHONG} and other properties of the background (see Refs.~\cite{Buric:2025anb,Buric:2025fye,Valach:2025saf,Afkhami-Jeddi:2025wra,Jia:2025jbi,Jia:2026pmv,Araya:2026shz,Giombi:2026kdz,Grozdanov:2026cut,Jia:2026ryl,Buric:2026qsp,Buric:2026cmn,Arnaudo:2026axe}). Intriguingly, such cavity QNM observables may even be able to diagnose (large) quantum effects on the black hole interior, which are expected to resolve the singularity.

In this paper, we have demonstrated that the measurement of $t_*$ may be, at least in principle, accessible to astrophysical measurements. The proposed setup seems to be best suited for probing small black holes, such as the hypothetical primordial black holes, in cases when they would be surrounded by thin transmissive accreting plasmas. As we showed, for certain parameter regimes, the transmitted EM radiation may lie within a frequency window accessible to radio astronomy, and allow for non-trivial transmission. Therefore, one may consider this setup as a potential future test for the existence of PBHs.  

For the moment, it appears that using cavity QNMs to probe larger black holes, e.g., Sgr A*, is technically impossible both due to the exponential suppression of the signal and its emitted low-frequency radiation. Future refinements of the setup should therefore be considered. Moreover, other future investigations should also focus on better understanding the effects of rotation, establish the existence of the cavity thermal product formula and the asymptotic QNM relations for Kerr black holes, better model the accretion disk and understand the signal interplay between radiation due to the black hole and the radiation emitted by the accretion disk. Finally, for probes of quantum effects, one may first wish to parametrise the effects of higher-derivative corrections on $t_*$ and other predictions discussed here. 


\begin{acknowledgments}
We thank Matej Bajec, Nejc Čeplak and Giuseppe Policastro for valuable discussions on related topics. The work of S.G.\ is supported by the STFC Ernest Rutherford Fellowship ST/T00388X/1. The work is also supported by the research programme P1-0402 and the project J7-60121 of Slovenian Research Agency (ARIS). V.M.\ is supported by the project J7-60121 of Slovenian Research Agency (ARIS). S.V.\ is supported by the Marie Skłodowska-Curie Actions programme GA-101177446 and the Slovenian Research and Innovation Agency (ARIS), contract number 5110-18/2025-5.
\end{acknowledgments}


\appendix

\section{The cavity thermal product formula and the asymptotic QNM spectra}\label{App:CTPF}

In this appendix, we derive the cavity thermal product formula and relate the cavity QNM spectrum to bouncing geodesics and the associated singularities of the retarded Green's function $G(t)$. We work in natural units. We consider a typical gauge-invariant single-channel fluctuation $\psi$ of some bulk field (be it scalar, electromagnetic or gravitational) that can be expressed in terms a (decoupled) master equation in Fourier space as (see \cite{Grozdanov:2023txs,Dodelson:2023vrw,Grozdanov:2026ktq})
\begin{equation}
\label{eq:Schrodinger FT}
    \left(\partial_z^2+\omega^2-V(z)\right)\psi(\omega,z)=0,
    \qquad 0\leq z<\infty,
\end{equation}
where the reflecting cavity wall lies at $z=0$, the black-hole horizon at $z\to\infty$, and $V(z)$ is real, frequency $\omega$-independent, regular at the cavity wall, and exponentially decaying near the horizon \cite{Grozdanov:2026ktq}.

Let $h_+(\omega,z)$ denote the solution ingoing at the horizon,
\begin{equation}
    h_+(\omega,z)\sim e^{ i\omega z},\qquad z\to\infty,
\end{equation}
and let $g(\omega,z)$ denote the solution satisfying $g(\omega,0)=0$, normalized by $\partial_z g(\omega,0)=1$. Then, the retarded Green function is
\begin{equation}
\label{eq: Green's Wronskian}
    G(\omega,z;z')
    = \frac{h_+(\omega,z')g(\omega,z)}{\mathcal{F}(\omega)}\,\theta(z'-z)
    + (z \leftrightarrow z'),
\end{equation}
where $\mathcal{F}(\omega)\equiv W[h_+,g]$ denotes the Wronskian of the two solutions. We assume thermal equilibrium at inverse temperature $\beta$, fixed by the black hole temperature. For the Schwarzschild-de Sitter black hole, this is allowed because the Dirichlet wall lies inside the static patch and excludes the cosmological horizon. The two-sided Wightman correlator is related to the retarded correlator by (see, e.g., \cite{Dodelson:2023vrw,Grozdanov:2025ulc})
\begin{equation}
\label{eq: G12 relation}
    G_{12}(\omega) = \frac{G(\omega)-G(-\omega)}
    {2i\sinh(\beta\omega/2)} .
\end{equation}
Using \eqref{eq: Green's Wronskian}, we then obtain
\begin{equation}
\label{eq: bulk G12}
    G_{12}(\omega,z;z') = \frac{\omega}{\sinh(\beta\omega/2)}
    \frac{g(\omega,z)g(\omega,z')}
    {\mathcal{F}(\omega)\mathcal{F}(-\omega)} .
\end{equation}
We further define a wall (boundary) correlator $G_{12}^{\partial}(\omega)$ at the timelike cavity wall (at $z=0$). Since the limit $z, z' \to 0$ of \eqref{eq: bulk G12} vanishes, the nontrivial boudary observable is obtained by taking normal derivatives at the wall. Equivalently, if
$h_+(\omega,z)=A(\omega)+B(\omega)z+\mathcal{O}(z^2)$ near $z=0$, then
\begin{equation}
    G^\partial(\omega)
    \propto
    \frac{B(\omega)}{A(\omega)}
    =\frac{\partial_z h_+(\omega,0)}{\mathcal{F}(\omega)} ,
\end{equation}
which gives
\begin{equation}
\label{eq: boundary G12}
    G_{12}^{\partial}(\omega)
    \propto
    \frac{\omega}{\sinh(\beta\omega/2)}
    \frac{1}{\mathcal{F}(\omega)\mathcal{F}(-\omega)} .
\end{equation}

The reflecting cavity permits a direct extension of the thermal product formula originally derived for holographic thermal correlators in AdS/CFT \cite{Dodelson:2023vrw}, to the wall correlators considered here. Crucially, no CFT input is required. The derivation uses only the analytic properties of the radial Schr\"odinger problem between the cavity wall and the black hole horizon, together with the asymptotics
\begin{equation}
\label{eq: potential asymp}
    V(z) \sim
    \begin{cases}
        \sum_{n=0}^{\infty} a_n z^n, & z \to 0 , \\[2mm]
        \sum_{n=0}^{\infty} b_n e^{-\frac{4 \pi n}{\beta}z}, & z \to \infty .
    \end{cases}
\end{equation}
The cavity setup is, in fact, simpler than the derivation in AdS/CFT because $V(z)$ is regular at the wall (at $z=0$). The factorisation then follows by applying the Weierstrass--Hadamard theorem to $1/G_{12}^{\partial}(\omega)$: an entire function of finite order admits a canonical product representation over its zeros, multiplied by the exponential of a polynomial \cite{book}. Thus, it suffices to establish that $1/G_{12}^{\partial}(\omega)$ is entire and of order one. 

The fact that $1/G_{12}^{\partial}(\omega)$ is entire follows from standard scattering theory results applied to the Wronskian $\mathcal F(\omega)$. The exponential horizon asymptotics in Eq.~\eqref{eq: potential asymp} imply that $\mathcal{F}(\omega)\mathcal{F}(-\omega)$ is meromorphic, with possible simple poles only at the nonzero Matsubara frequencies (see, e.g., Ref.~\cite{Newton:1982qc}),
\begin{equation}
     \omega=\frac{2\pi i n}{\beta},
    \qquad n\in\mathbb{Z}\setminus\{0\}.
\end{equation}
In Eq.~\eqref{eq: boundary G12}, these poles are exactly canceled by the zeros of $\sinh(\beta\omega/2)/\omega$, and, hence, $1/G_{12}^{\partial}(\omega)$ is entire.

To determine the growth order, we use the large-$|\omega|$ behaviour of $\mathcal F(\omega)$. Since the cavity potential is regular at $z=0$, scattering theory estimates apply directly \cite{Newton:1982qc}. In particular, if
\begin{equation}
\label{eq: cond z}
    \int_0^\infty dz\, z |V(z)| < \infty ,
\end{equation}
then $\mathcal F(\omega)\to 1$ as $|\omega|\to\infty$ for $\Im\omega\geq 0$. Assuming $V(z)$ is analytic, this property extends into the lower half-$\omega$-plane away from the line of the Matsubara poles \cite{Newton:1982qc}. Thus,
\begin{equation}
\label{eq: G12 asymptotics}
    \frac{1}{G_{12}^{\partial}(\omega)}
    \propto
    \frac{\sinh(\beta\omega/2)}{\omega}
    \mathcal F(\omega)\mathcal F(-\omega)
    \sim
    \frac{1}{\omega}e^{\beta\omega/2},
\end{equation}
along rays asymptotically avoiding the poles. The remaining direction is the imaginary $\omega$ axis. In the AdS-Schwarzschild case, Ref.~\cite{festucciathesis} showed that the large-imaginary-$\omega$ behaviour of the two-sided Wightman correlator is governed by the bouncing geodesic,
\begin{equation}
\label{eq: i omega asymptotics}
    G_{12}^{\partial}(\omega)
    \sim e^{\pm i\omega \Re(t_*)},
    \qquad
    \omega\to \pm i\infty ,
\end{equation}
up to polynomial prefactors, where $t_*$ is the complex bouncing time. Since the same structural inputs are present in the cavity problem, namely the exponential horizon asymptotics and the existence of a bouncing geodesic, we assume that the same large-imaginary-$\omega$ asymptotics apply. Under this assumption, $1/G_{12}^{\partial}(\omega)$ has at most exponential growth in all directions, and, hence, is of order one.

Finally, the symmetries of the radial problem imply that generic poles of $G_{12}^{\partial}(\omega)$ occur in quartets,
\begin{equation*}
     (\omega_n,-\omega_n,\omega_n^*,-\omega_n^*) .
\end{equation*}
Applying the Weierstrass--Hadamard factorisation theorem to $1/G_{12}^{\partial}(\omega)$ then yields the cavity thermal product formula:
\begin{equation}
    G_{12}^{\partial}(\omega)
    =
    \frac{G_{12}^{\partial}(0)}
    {\displaystyle\prod_{n=1}^\infty
    \left(1-\frac{\omega^2}{\omega_n^2}\right)
    \left(1-\frac{\omega^2}{(\omega_n^*)^2}\right)} .
\end{equation}
An analogous formula holds for the bulk two-sided correlator \eqref{eq: bulk G12}, with an additional numerator product over the zeros $\tilde{\omega}_n(z)$ of $g(\omega,z)$.

This product formula also fixes the large-overtone structure of the cavity QNM spectrum. In Ref.~\cite{Dodelson:2023vrw}, the analogous step used the CFT operator product expansion (OPE) asymptotics at large real frequency. Here, this input is replaced by the scattering theory asymptotics \eqref{eq: G12 asymptotics}, which have the same leading large-frequency form as the OPE asymptotics after setting $2\Delta-d=1$. Thus, the argument of Refs.~\cite{Dodelson:2023vrw,Dodelson:2025jff} can be translated directly to the present cavity problem.

Assume that the upper half-plane representatives of the high-$n$ poles
asymptotically lie on a single ray,
\begin{equation}
\label{eq: subleading ansatz}
    \hat \omega_n = r e^{i\theta} n + s e^{i\phi}+\cdots,\quad n\to\infty.
\end{equation}
Inserting this ansatz into Eq.~\eqref{eq: CTPF} and matching to Eq.~\eqref{eq: G12 asymptotics} gives the leading and first subleading constraints
\begin{equation}
    \beta=\frac{4\pi\sin\theta}{r},\quad
    \frac{4s\cos(\theta-\phi)+2r}{r} = 1.
\end{equation}
Moreover, Fourier transforming the thermal product formula, one finds \cite{Dodelson:2025jff} that $G^\partial_{12}(t)$ has singularities at
\begin{equation}
\label{eq: hattnm}
    \hat t_{nm} = \frac{i\beta}{2} + 2\pi\left(\frac{n}{\Omega^*}  - \frac{m}{\Omega} \right), \quad n,m\in\mathbb Z_{\geq0},
\end{equation}
together with the reflected points $-\hat t_{nm}$. Here, $2\Delta-d=1$, appropriate to our case, and $\Omega\equiv re^{i\theta}$, so the leading behaviour is  $\hat\omega_n\sim\Omega n$. The corresponding retarded cavity QNMs
satisfy $\omega_n\sim\Omega^* n$.

Although the retarded Green's function contains a Heaviside factor and therefore has no single natural analytic continuation in time, its positive- and negative-time branches can be analytically continued separately. For details, see Appendix B of Ref.~\cite{Grozdanov:2026cut}. The lattice \eqref{eq: hattnm} then gives the retarded singularities:
\begin{equation}
\label{eq: tnm}
    t_{nm}=2\pi\left(\frac{n}{\Omega^*}-\frac{m}{\Omega}\right),
    \quad n,m\in\mathbb Z .
\end{equation}
The first nontrivial points in the positive- and negative-time branches, $t_{10}$ and $t_{01}$, correspond, respectively, to the bouncing time $t_*$ and its reflected counterpart \cite{Dodelson:2025jff,Grozdanov:2026cut}. Indeed, by using the local Hadamard form together with propagation of singularities theorem \cite{Grozdanov:2026ktq}, singularities of $G(t)$ propagate along null geodesics; for coincident spatial insertions, the first nontrivial such geodesic is the bouncing geodesic. Hence, $t_*=t_{10}$, and we arrive at the final (universal) expression for the asymptotic behaviour of black hole QNMs inside a reflecting cavity with any value of the cosmological constant:
\begin{equation}
    \omega_n\sim\frac{2\pi n}{t_*},\quad n\to\infty .
\end{equation}


\section{Transmission coefficient through an idealised transparent plasma}\label{a.WKBB}

To compute the transmission coefficient of electromagnetic waves passing through an idealised model of an accreting plasma, we consider only the $\gamma=0$ case of the master equation \eqref{e.apple_on_tree} with the potential \eqref{e.rozok}. Moreover, we will only study a special limit of $W(r)$ approximated by a step function in Eq.~\eqref{e.rectangle_is_just_a_sexy_square} and neglect the angular-momentum-dependent term $\ell(\ell+1)/r^2$ because it is subleading in the plasma-dominated regime assumed below.

We first define $r_*^{(1)}\equiv r_*(r_a)$ and $r_*^{(2)}\equiv r_*(r_a+L)$, where the tortoise coordinate reads
\begin{equation}
    r_*=r+r_b\ln\left(\frac{r}{r_b}-1\right).
\end{equation}
We also define effective wave numbers $k_0$ and $k(r_*)$ by
\begin{equation}
    k_0\equiv\frac{\omega}{c},\,\,\,\,\,k(r_*)\equiv\sqrt{\frac{\omega_p^2}{c^2}\left(1-\frac{r_b}{r(r_*)}\right)-\frac{\omega^2}{c^2}},
\end{equation}
divide the system into three regions: $r_*<r_*^{(1)}$ (region I), $r_*^{(1)}<r_*<r_*^{(2)}$ (region II) and $r_*>r_*^{(2)}$ (region III). The solutions to \eqref{e.apple_on_tree} in these regions read
\begin{align}
    \psi_{\rm I}(r_*)&=Ae^{ik_0(r_*-r_*^{(1)})}+Be^{-ik_0(r_*-r_*^{(1)})},\\
    \psi_{\rm II}(r_*)&\approx \frac{F}{\sqrt{k(r_*)}}e^{\mathcal{S}(r_*)}+\frac{G}{\sqrt{k(r_*)}}e^{-\mathcal{S}(r_*)},\\
    \psi_{\rm III}(r_*)&=Ce^{ik_0(r_*-r_*^{(2)})},   
\end{align}
where $\mathcal{S}(r_*)=\int_{r_*^{(1)}}^{r_*}k(r'_*)\dd r'_*$, in region III we assumed only an outgoing wave, and inside the barrier (region II), we used the standard WKB ansatz. 

Requiring continuity of $\psi$ and its first derivative at $r_*^{(1)}$ and $r_*^{(2)}$, and using the large-$\omega_p$ limit, we can express the coefficient $A$ via $C$,
\begin{equation}\label{e.pigeon_bbq}
    A\approx\frac{-C\sqrt{k^{(1)}k^{(2)}}}{2ik_0}\sinh{\mathcal{S}(r_*^{(2)})},
\end{equation}
where $k^{(1)}=k(r_*^{(1)})$, $k^{(2)}=k(r_*^{(2)})$, and
\begin{equation}\label{e.rabbit_soup}
\begin{split}
    \mathcal{S}(r_*^{(2)})=&\int_{r_*^{(1)}}^{r_*^{(2)}}k(r_*)\dd r_*=\int_{r_a}^{r_a+L}k(r)\dv{r_*}{r}\dd r\\
    \approx&\frac{\omega_p}{c}\int_{r_a}^{r_a+L}\sqrt{\frac{r}{r-r_b}}\dd r=\frac{\omega_p}{c}\Big[ \sqrt{r(r-r_b)}\\
    &+r_b \ln\left(\sqrt{r} + \sqrt{r-r_b}\right) \Big]_{r_a}^{r_a+L}.
\end{split}
\end{equation}
Finally, the transmission coefficient $T$ is given by $\abs{C/A}^2$, which, using Eqs.~\eqref{e.pigeon_bbq} and \eqref{e.rabbit_soup}, yields the final result stated in Eqs.~\eqref{e.hamster} and \eqref{e.hamsters_friend}.


\bibliography{draft}
\bibliographystyle{apsrev4-1}

\end{document}